# Can ChatGPT assist visually impaired people with micro-navigation?

Junxian He [1], Shrinivas Pundlik [2], Gang Luo [2]

1. Chongqing University, Chongqing, China

2. Schepens Eye Research Institute of Mass Eye & Ear, Harvard Medical School, Boston MA

Corresponding Author:

Gang Luo (gang.luo@schepens.harvard.edu)

**Abstract**

Objective
Micro-navigation poses challenges for blind and visually impaired individuals. They often need to ask for sighted assistance. We explored the feasibility of utilizing ChatGPT as a virtual assistant to provide navigation directions.

Methods
We created a test set of outdoor and indoor micro-navigation scenarios consisting of 113 scene images and their human-generated text descriptions. A total of 412 way-finding queries and their expected responses were compiled based on the scenarios. Not all queries are answerable based on the information available in the scene image. "I do not know" response was expected for unanswerable queries, which served as negative cases. High level orientation responses were expected, and step-by-step guidance was not required. ChatGPT 4o was evaluated based on sensitivity (SEN) and specificity (SPE) under different conditions.

Results
The default ChatGPT 4o, with scene images as inputs, resulted in SEN and SPE values of 64.8% and 75.92%, respectively. Instruction on how to respond to unanswerable questions did not improve SEN substantially but SPE increased by around 14 percentage points. SEN and SPE both improved substantially, by about 17 and 16 percentage points on average respectively, when human written descriptions of the scenes were provided as input instead of images. Providing further prompt instructions to the assistants when the input was text description did not substantially change the SEN and SPE values.

Conclusion
Current native ChatGPT 4o is still unable to provide correct micro-navigation guidance in some cases, probably because its scene understanding is not optimized for navigation purposes. If multi-modal chatbots could interpret scenes with a level of clarity comparable to humans, and also guided by appropriate prompts, they may have the potential to provide assistance to visually impaired for micro-navigation.

**Introduction**

Blind and visually impaired individuals (BVI) often face challenges related to orientation and mobility in their daily-life tasks. Navigation, or wayfinding, is one of the key components of mobility. Assistive devices and technologies, including mobile apps, are intended to help BVI individuals with a wide-variety of daily-life tasks, including navigation.[1] Navigation can be further classified as macro- or micro-navigation. If navigation is considered as the task of getting from point A to point B in its entirety, macro-navigation refers to the high-level aspects of path/route planning and following-up, generally over larger distances (say over many miles), and often facilitated by geo-localization/mapping technologies.[2,3] Micro-navigation, on the other hand, refers to navigating within a close range during the journey – precisely to a particular location, for instance, finding store entrances, train station exits, and so on. Micro-navigation is often ignored or taken for granted in the general context of navigation. While a macro-navigation tool can guide BVI individuals to the vicinity of a building, getting into the building, or getting to the elevator entrance, can be challenging micro-navigation tasks that are essential to the entire trip. However, this is exactly the kind of the task that is largely unresolved for BVI individuals who want to travel independently,[4,5] as there is no general-purpose micro-navigation tool or aid and they frequently have to resort to asking others for help.

While there are a large number of devices and smartphone apps for navigation assistance, overwhelming number of them are for macro-navigation.[2,3] Mapping and geo-location based apps are not particularly suitable or are inadequate for micro-navigation, not only because of errors in geo-localization and mapping,[6] but also because of lack of mapping as it is not feasible to accurately map all the locations. Some vision aids and assistance apps make use of computer vision algorithms to perform object detection[1] or provide micro-navigation assistance in certain specific scenarios – such as public transit specific information.[7,8] However, the sheer variety of objects one could encounter in the real-world makes using custom tool for specific objects somewhat infeasible. Apps and services such as Aira, which provide live, remote personalized assistance to BVI individuals could be helpful in navigation.[9-12] However, cost and feasibility of human assistance means that its utilization tends to be limited. A 24×7 virtual assistant could alleviate many of these micro-navigation related challenges for BVI individuals. Such a virtual assistant for BVI travelers is not out of the realm of possibility, given the recent advances in artificial intelligence models for computer vision, large language models (LLMs), and visual-language models.[13,14]

Visual-language models or vision-language models (VLMs) are designed for tasks that require some combination of computer vision and natural-language inputs/outputs such as image captioning or visual question answering, among others.[13] Multi-modal VLMs have been employed for navigation (vision-language navigation or VLN), especially for robot navigation, where the idea is that the robot will take natural language instructions, extract salient information from the text (such as landmarks and their inter-relationship), and then perform the navigation tasks based on visual detection of the said landmarks in previously unseen environments.[15-25] While the VLM approaches are focused understanding image- and language-based cues for navigation by autonomous agents, our goal is slightly different: providing direct and precise responses to navigation-related queries of human users. The scene understanding component by extracting navigation-relevant information from images is common in our approach, however, instead of providing instructions to the agent, the agent it required to generate specific instructions for humans based on natural language interaction.

Abilities of AI models to provide description of a scene and interact with humans is now evident, for example GPT-4 with image inputs[26] or a large language-and-vision assistant (LLVA).[19] We evaluated different approaches of training an LLM with the goal of providing accurate navigation directions based on a given image input of a scene. We generated text description of scene images, along with positive and negative query-response pairs corresponding to each scene as the base training dataset. From there on, we tried various kinds of fine tuning and prompt-based training approaches with the goal of eliciting actionable responses from the LLM. Our hypothesis was that with sufficient training data, a contemporary LLM could be trained to provide accurate, on-demand, scenario-specific navigation information that could be useful to people with vision loss.

**Methods**

*Dataset Generation*

The dataset consisted of real-world 113 images of indoor and outdoor locations representative of typical navigation scenarios along with human generated, textual descriptions of the scene depicted in these images (Figure 1). The outdoor scenarios included street/sidewalks, plaza, transit stations, office campus, whereas, the indoor scenarios consisted of supermarket aisle, shopping malls interiors, and subway stations. The images were captured with mobile phone cameras with the perspective of the potential user who would like to query information. Short descriptions of the captured picture, typically a few lines, were written by members of the study team. The description included the general characteristics of the scene as well as specific inter-relationships of the various objects present in the scene. Emphasis was provided to make the description informative and relevant to navigation-related queries that may potentially arise. Additionally, 4 navigation-related queries were generated by the humans for each image/scenario: 2 answerable and 2 unanswerable. Answerable queries were those related to the information present in the scene image and the corresponding text description. Thus, a human looking at the picture and/or scene text could reasonably answer the question and provide specific navigation related guidance. On the other hand, unanswerable queries were those for which information was not present and the expected answer was "I do not know". From 113 images and their associated text descriptions, 412 total queries and their template responses were created. All of this data was human generated.

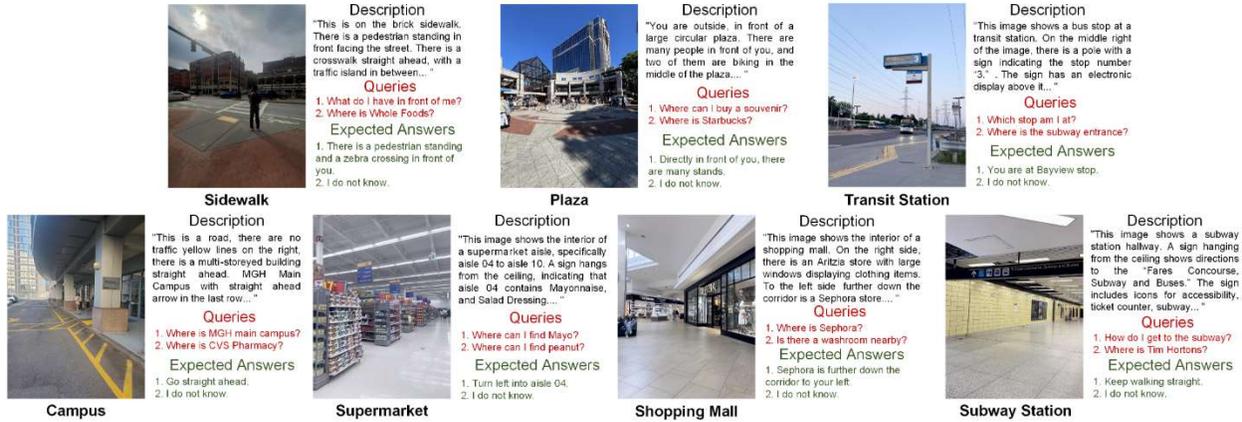

**Figure 1. The dataset of navigation scenarios consisted of 113 indoor and outdoor images along with their human generated text description of the scene, navigation-related queries based on the scene (positive and negative), and the expected answers to the queries.**

*Virtual Micro-navigation Assistants*

We created 3 different versions of micro-navigation assistants based on OpenAI GPT4-o model with: i) default system instructions (means just create an assistant but do not edit or change any settings), ii) simple system instructions, and iii) advanced system instructions (Table 1). These instructions can be provided via python code or via OpenAI assistant UI (see Appendix I). Each of the 3 versions were fed image or text description as input, along with the 4 navigation-related queries. Responses from the virtual assistants to each query was collected and evaluated for their accuracy in providing navigation guidance. The memory of the assistants' being evaluated was cleared between successive image inputs to avoid memorization.

*Performance Evaluation*

The ability of the virtual assistants to provide navigation assistance was evaluated based on their sensitivity and specificity from the obtained responses. Sensitivity (SEN) was computed as the proportion of true positive to actual positive cases (answerable queries). Specificity (SPE) was the proportion of true negative to actual negative (unanswerable queries). In the case of answerable queries, true positive responses were those that could be considered as consistent with the corresponding human generated responses. Those inconsistent with human generated responses (incorrect or made-up responses) were considered as false negatives. In the case of unanswerable queries (when the sought after information is not present in the scene image and associated textual description), the response "I do not know" or something like that was considered as true negative, whereas any other response (made-up) was considered as false positives. In order to judge the similarity of responses to the queries from the navigation assistants and the human generated ground truth, we used a separate GPT4-o model-based grading assistant (Table 2) to compare the two responses and provide a "yes" (similar answers or a match) or "no" (different answers) verdict that were used to calculate SEN and SPE rates for various conditions. The grading assistant itself was evaluated by determining the degree of agreement between its judgment (whether ground truth and GPT responses are similar) and human judgment in a subset of queries (n=132). Cohen's Kappa coefficient was computed to determine the inter-rater reliability (between auto-grading and human grading), with the expectation that there would a high-level of consistency between the judgment of auto grader and human grader, thereby indicating the feasibility of using GPT4-o to determine the accuracy of the navigation responses.

**Table 1. Instructions provided to the LLM micro-navigation assistants.**

| Assistant | Instruction |
|---|---|
| Default-Navigation | N/A |
| Simple-Navigation | You are a navigator that should give correct directions to help a person go to where they need to go. |

| Advanced -Navigation | You are a navigator specifically made to help people with vision disabilities or blind people. Your task is to suggest a walking direction to respond to a navigation question in the user message. Before you answer, please think step by step. Your answer should be correct. If the information provided is not enough to form a full set of directions to the destination, do not output any directions. |

Table 2. The instructions provided to the LLM grading assistant.

| Assistant | Instruction |
| --- | --- |
| Grading | You will receive two sets of directions for a location, each in parenthesis and brackets. Your job is to determine whether they are roughly consistent by saying "Yes" or "No". The directions don't have to match word for word; the focus is on whether they lead to the same general direction or not. |

**Results**

*Reliability of automated grading of responses*

Of the 132 queries used for evaluating the agreement between GPT4-o grading assistant and human grader, 22 were negative (16.7%) and the rest were positive. There was an agreement between the grading assistant and human grader on 113 responses (86%) – 100% agreement for negative queries and 83% on positive queries. The inter-rater reliability between the two methods was high (Cohen's κ = 0.71, z=8.21, p<0.001), indicating that auto grading via GPT4-o grading assistant could be reliably used for inferring whether the responses provided by navigation assistants were consisted with the human generated ground truth responses to navigation queries.

There was no significant effect of ChatGPT's response length (number of words) on the odds of agreement with the ground truth response.

*Effectiveness of navigation assistants*

The SPE and SEN the 3 navigation assistants in two different conditions (text descriptions or image) are shown in Table 3. Default configuration (without any additional instructions) with image as input resulted in the worst SEN and SPE values. With advanced instructions, SEN with image input did not improve substantially but SPE increased by around 14 percentage points. SEN and SPE both improved substantially, by about 17 and 16 percentage points on average respectively, when text descriptions of the scenes were provided as input instead of images. Providing further prompt-instructions to the assistants when the input was text description did not substantially change the SEN and SPE values.

Table 3. The SEN and SPE of assistants' responses

| Provided Data | Assistant | SEN | SPE |
|---|---|---|---|
| Image | Default-Navigation | 64.8% | 75.92% |
| | Simple-Navigation | 63.6% | 85.19% |
| | Advanced -Navigation | 65.2% | 89.51% |
| Text Description | Default-Navigation | 80.8% | 99.38% |
| | Simple-Navigation | 83.6% | 98.76% |
| | Advanced -Navigation | 79.2% | 99.38% |

.

**Discussion**

In this study, we investigated the ability of AI chat bots to understand visual scenes and deliver natural language navigation instructions to BVI individuals. By collecting a dataset composed of pictures of a variety of real-world navigation scenes, their corresponding text descriptions, navigation-related queries associated with the scenes and their ground truth responses, we were able to compare the efficacy of the various combinations of inputs and tasks instructions to ChatGPT4o in answering navigation-related queries of BVI individuals. In general, the SEN and SPE of the navigation assistants were higher when provided with human written description of scenes instead of image input. However, even with text as input, the SEN on average was around 83%. The overall results indicate that the current general-purpose LLMs and vision-language models are not reliable enough for working as micro-navigation assistants to BVI individuals. Providing instructions to the assistants did lead to substantial improvement in the case of text input, although specificity improved with advanced instructions together with image input. This indicates that even simple prompt-based instructions, can help in answering negative queries by reducing hallucination.

There are two major reasons for lower SEN of navigation assistants, especially with image inputs. First, the navigation assistants could not interpret images when the queried destination is related to directional signage or if it does not occupy large enough area in the scene image. The LLM sometimes cannot recognize the directional signage correctly. For example, failing to recognize left pointing arrow, the LLM tells the user to go straight (Figure 2). Directional signage typically appear small in size in scene images, which could further complicate the issue as the LLM is unable to recognize small targets. Second, even though the LLM recognized the destination or a target in the image, it sometimes gave vague answers without sufficient details or further directions about how to specifically go to the destination (Figure 3). Lower SPE is mainly caused by the hallucination of LLM and providing instructions to the assistant helps alleviate this issue to some extent.

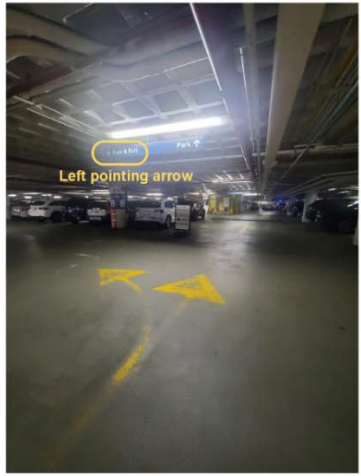

Figure 2. The LLM failing to recognize the left pointing arrow in the image and tells the user to go straight.

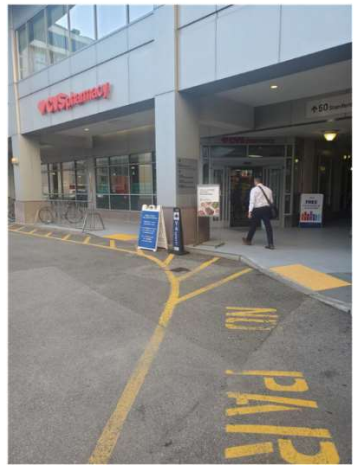

Figure 3. The LLM failing to give specific directions even though it recognizes the destination or a target in the image.

Although the SEN is better with human written descriptions as input compared to image, it is still not high enough because the LLM navigation assistants sometimes combine logically irrelevant but neighboring sentences as one destination description, resulting in incorrect directions to the user. Hallucination phenomenon is also seen with text as input because the LLM considers the queried destination as related to something mentioned in text descriptions, although there is no relationship between them.

BVI individuals could potentially benefit from advances in AI and LLMs. A recent survey reported that ChatGPT has been rapidly adopted by BVI users in their daily life tasks, and the majority of those who were aware of it but not a frequent users were interested in learning more about it.[27] In the context of navigation, BVI users require precise instructions and ability to improvise in unexpected situations. Till date, this kind of task was only possible in the realm of human agents. But if AI virtual assistants could prove to be adequate, then accessibility and quality of life of BVI could be improved. In order to do that, navigation assistants need to be improved. The navigation assistants could potentially be beneficial for BVI micro-navigation if the images are well-taken and if proper instructions are given to the LLM during execution. The performance could be improved by by training custom vision-language models for navigation-specific scenarios involving BVI travelers. Future work involves evaluating the utility of navigation directions provided by navigation assistants by BVI users.

**Conclusion**

Current native ChatGPT 4o is still unable to provide correct micro-navigation guidance in some cases, probably because its scene understanding is not optimized for navigation purposes. If multi-modal chatbots could interpret scenes with a level of clarity comparable to humans, and also guided by appropriate prompts, they may have the potential to provide assistance to visually impaired for micro-navigation.

4
**Acknowledgements**

We thank Prerana Shivshanker, Tony Liu, Jessie Zhou and Joye Luo for their assistance in creating testing dataset and editing the paper.

**References**

1. Pundlik S, Shivshanker P, Luo G. Impact of Apps as Assistive Devices for Visually Impaired Persons. *Annual Reviews of Vision Science*. 2023;9:12.11-12.20.
2. Parker AT, Swobodzinski M, Wright JD, Hansen K, Morton B, Schaller E. Wayfinding Tools for People With Visual Impairments in Real-World Settings: A Literature Review of Recent Studies. *Frontiers in Education*. 2021;6.
3. Swobodzinski M, Parker AT. *A Comprehensive Examination of Electronic Wayfinding Technology for Visually Impaired Travelers in an Urban Environment: Final Report. NITC-RR-1177*. Portland, OR: Transportation Research and Education Center (TREC);2019.
4. Crudden A. Transportation and Vision Loss: Where are we Now? . *The Journal of American Society of Opthalmic Registered Nurses*. 2018;43(2):19-24.
5. Crudden A, McDonnall MC, Hierholzer A. Transportation: An Electronic Survey of Persons who Are Blind or Have Low Vision. *Journal of Visual Impairment & Blindness*. 2015;109(6):445-456.
6. Luo G, Pundlik S. Widespread Errors in Bus Stop Location Mapping is an Accessibility Barrier for Passengers Who are Blind or Have Low Vision. *Journal of Visual Impairment & Blindness*. 2023.
7. Feng J, Beheshti M, Philipson M, Ramsaywack Y, Porfiri M, Rizzo JR. Commute Booster: A Mobile Application for First/Last Mile and Middle Mile Navigation Support for People with Blindness and Low Vision. *IEEE Journal of Translational Engineering in Health and Medicine*. 2023:1-1.
8. Pundlik S, Shivshanker P, Traut-Savino T, Luo G. Field Evaluation of a Mobile App for Assisting Blind and Visually Impaired Travelers to Find Bus Stops. *Translational Vision Science & Technology*. 2024;13(1):11-11.
9. Aira. https://aira.io/.
10. Nguyen, B J, Chen WS, Chen AJ, et al. Large-scale assessment of needs in low vision individuals using the Aira assistive technology. *Clin Ophthalmol*. 2019;13:1853-1868.
11. Nguyen BJ, Kim Y, Park K, et al. Improvement in patient-reported quality of life outcomes in severely visually impaired individuals using the aira assistive technology system. *Translational vision science & technology*. 2018;7(5):30-30.
12. Park K, Kim Y, Nguyen BJ, Chen A, Chao DL. Quality of Life Assessment of Severely Visually Impaired Individuals Using Aira Assistive Technology System. *Trans Vis Sci Tech*. 2020;9(4):21.

**Appendix I**

Code (1)

```
from openai import OpenAI
client = OpenAI()
assistant = client.beta.assistants.create(
 name=" Simple-Navigation",
 instructions=" You are a navigator that should give correct directions to help a person go to where they need to go.",
model="gpt-4o",
)
```

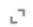

Figure A1. Ways of providing instructions to ChatGPT bots: via code (top) or by prompt.